\documentclass[10pt, conference]{IEEEtran}
\usepackage{subfigure}
\usepackage{epstopdf}
\usepackage{epsfig}
\usepackage{caption}
\usepackage{array}
\usepackage{lipsum}
\usepackage[]{algorithm2e}
\usepackage{tabularx}
\usepackage[bookmarks=false]{hyperref}

\usepackage{booktabs} 

\usepackage{graphicx}
\usepackage{booktabs}
\usepackage{float}
\usepackage{multirow}
\usepackage{multicol}
\usepackage{array}
\usepackage{tabularx}
\usepackage{bm}
\usepackage{soul}
\usepackage{tikz}
\usepackage{amssymb}
\usepackage{amsmath}
\usepackage{xfrac}
\usepackage[bottom]{footmisc}
\usepackage{fixltx2e}

\usepackage{etoolbox}

\renewcommand{\IEEEbibitemsep}{0pt plus 0.5pt}
\makeatletter
\IEEEtriggercmd{\reset@font\normalfont\fontsize{7.0pt}{8.0pt}\selectfont}
\makeatother
\IEEEtriggeratref{1}

\begin{document}
\fontsize{9.5}{11.5}\selectfont
\title{Workload-Aware Opportunistic Energy Efficiency in Multi-FPGA Platforms}


\author{Sahand Salamat, Behnam Khaleghi, Mohsen Imani, and Tajana Rosing\\
Computer Science and Engineering Department, UC San Diego, La Jolla, CA 92093, USA\\
    \{sasalama, bkhaleghi, moimani, tajana\}@ucsd.edu }
\maketitle

\begin{abstract}
The continuous growth of big data applications with high computational and scalability demands has resulted in increasing popularity of cloud computing.
Optimizing the performance and power consumption of cloud resources is therefore crucial to relieve the costs of data centers.
In recent years, multi-FPGA platforms have gained traction in data centers as low-cost yet high-performance solutions particularly as acceleration engines, thanks to the high degree of parallelism they provide.
Nonetheless, the size of data centers workloads varies during service time, leading to significant underutilization of computing resources while consuming a large amount of power, which turns out as a key factor of data center inefficiency, regardless of the underlying hardware structure.
In this paper, we propose an efficient framework to throttle the power consumption of multi-FPGA platforms by dynamically scaling the voltage and hereby frequency during runtime according to \emph{prediction} of, and \emph{adjustment} to the workload level, while maintaining the desired Quality of Service (QoS).
This is in contrast to, and more efficient than, conventional approaches that merely scale (i.e., power-gate) the computing nodes or frequency.
The proposed framework carefully exploits a pre-characterized library of delay-voltage, and power-voltage information of FPGA resources, which we show is indispensable to obtain the efficient operating point due to the different sensitivity of resources w.r.t. voltage scaling, particularly considering multiple power rails residing in these devices.
Our evaluations by implementing state-of-the-art deep neural network accelerators revealed that, providing an average power reduction of 4.0$\times$, the proposed framework surpasses the previous works by 33.6\% (up to 83\%).

\end{abstract}

\section{Introduction} \label{sec:intro}
The emergence and prevalence of big data and related analysis methods, e.g., machine learning on the one hand, and the demand for a cost-efficient, fast, and scalable computing platform on the other hand, have resulted in an ever-increasing popularity of cloud services where services of the majority of large businesses nowadays rely on cloud resources \cite{bhattacherjee2014end}.
In a highly competitive market, cloud infrastructure providers such as Amazon Web Services, Microsoft Azure, and Google Compute Engine, offer high computational power with affordable price, releasing individual users and corporations from setting up and updating hardware and software infrastructures.
Increase of cloud computation demand and capability results in growing hyperscale cloud servers that consume a huge amount of energy.
In 2010, data centers accounted for 1.1-1.5\% of the world's total electricity consumption \cite{wahlroos2018future}, with a spike to 4\% in 2014 \cite{shehabi2016united} raised by the move of localized computing to cloud facilities.
It is anticipated that the energy consumption of data centers will double every five years \cite{tseng2018energy}. 

The huge power consumption of cloud data centers has several adverse consequences \cite{altomare2018data}: (i) operational cost of cloud servers obliges the providers to rise the price of services, (ii) high power consumption increases the working temperature which leads to a significant reduction in the system reliability as well as the data center lifetime, and (iii) producing the energy required for cloud servers emits an enormous amount of environmentally hostile carbon dioxide. Therefore, improving the power efficiency of cloud servers is a critical obligation.

That being said, several specialized hardware accelerators \cite{magaki2016asic, salamat2018rnsnet} or ASIC-ish solutions \cite{jouppi2017datacenter, chen2014diannao} have been developed to increase the performance per watt efficiency of data centers.
Unfortunately, they are limited to a specific subset of applications while the applications and/or implementation of data centers evolve with a high pace. 
Thanks to their relatively lower power consumption, fine-grained parallelism, and programmability, in the last few years, \emph{Field-Programmable Gate Arrays} (FPGAs) have shown great performance in various applications\cite{salamat2019f5, imani2019fach, imani2019sparsehd, mahajan2016tabla, sharma2016high}. Therefore, they have been integrated in data centers to accelerate the data center applications.
Cloud service providers offer FPGAs as Infrastructure as a Service (IaaS) or use them to provide Software as a Service (SaaS).
Amazon and Azure provide multi-FPGA platforms for cloud users to implement their own applications.
Microsoft and Google are other big names of corporations/companies that also provide applications as a services, e.g., convolutional neural networks \cite{ovtcharov2015accelerating}, search engines \cite{putnam2014reconfigurable}, text analysis \cite{weerasinghe2016network}, etc. using multi-FPGA platforms.


Having all the benefits blessed by FPGAs, underutilization of computing resources is still the main contributor to energy loss in data centers.
Data centers are expected to provide the required QoS of users while the size of the incoming workload varies temporally.
Typically, the size of the workload is less than 30\% of the users' expected maximum, directly translating to the fact that servers run at less than 30\% of their maximum capacity \cite{altomare2018data}.
Several works have attempted to tackle the underutilization in FPGA clouds by leveraging the concept of \textit{virtual machines} to minimize the amount of required resources and turn off the unused resources \cite{zhang2018survey}.
In these approaches, FPGA floorplan is split into smaller chunks, called virtual FPGAs, each of which hosts a virtual machine.
FPGA virtualization, however, degrades the performance of applications, congests routing, and more importantly, limits the area of applications \cite{yazdanshenas2017quantifying}.
This scheme also suffers from security challenges \cite{yazdanshenas2018interconnect}.

A straightforward technique to get around with the underutilization of computing nodes is to adjust the operating frequency in tandem with workload variation where all nodes are still responsible for processing a portion of input data.
This reduces the dynamic power consumption proportional to the workload, and resolves the problem of wake-up and reconfiguration time that come with power gating of nodes.
Nonetheless, as all nodes are active, the static power remains a challenge especially in elevated temperatures near FPGA boards in data centers \cite{putnam2014reconfigurable} that exponentially increase the leakage current.
Dynamic voltage \emph{and} frequency scaling (DVFS) is a promising technique to resolve this problem by scaling the voltage according to the available performance headroom.
That is, the circuit does not need to use the nominal voltage when it is not required to deliver the maximum performance \cite{chow2005dynamic}.

\emph{Optimum} voltage and frequency scaling in FPGAs, however, is sophisticated.
The critical path in FPGA-based designs is application-dependent.
Therefore, employing prefabricated representative critical path sensors (e.g., ring oscillators) to examine the timing of designs as done for ASICs and processors is not practical \cite{drake2007distributed}.
Moreover, FPGAs comprise a heterogeneous set of components, e.g., logic look-up tables (LUTs), interconnection resources, DSPs, on-chip block RAMs and I/Os with separate voltage rails.
Obtaining the optimal point of voltages that minimizes the power while meets the (scaled) performance constraint is challenging.
As we investigate later in this paper, optimal operating voltages depend on critical path(s) resources, utilized (i.e., application) resources and their activity as well as total available resources, and the workload.

The main focus of this work is optimizing the energy consumption of multi-FPGA data center platforms, accounting for the fact that the workload is often considerably less than the maximum anticipated.
We leverage this opportunity to still use the available resources while efficiently scale the voltage of the entire system such that the projected throughput (i.e., QoS) is delivered.
We utilize a light-weight predictor for proactive estimation of the incoming workload and incorporate it to our power-aware timing analysis framework that adjusts the frequency and finds \emph{optimal} voltages, keeping the process transparent to users.
Analytically and empirically, we show the proposed technique is significantly more efficient than conventional power-gating approaches and memory/core voltage scaling techniques that merely check timing closure, overlooking the attributes of implemented application.

\section{Related Work} \label{sec:related}

The use of FPGAs in modern data centers have been gained attention recently as a response to rapid evolution pace of data center services in tandem with the inflexibility of application-specific accelerators and unaffordable power requirement of GPUs \cite{yazdanshenas2017quantifying, weerasinghe2016network}.
Data center FPGAs are offered in various ways, Infrastructure as a Service for FPGA rental, Platform as a Service to offer acceleration services, and Software as a service to offer accelerated vendor services/software \cite{weerasinghe2015enabling}.
Though primary works deploy FPGAs as tightly-coupled server addendum, recent works provision FPGAs as an ordinary standalone network-connected server-class node with memory, computation and networking capabilities \cite{weerasinghe2015enabling, weerasinghe2016network}.
Various ways of utilizing FPGA devices in data centers have been well elaborated in \cite{yazdanshenas2017quantifying}.

FPGA data centers, in parts, address the problem of programmability with comparatively less power consumption than GPUs.
Nonetheless, the significant resource underutilization in non-peak workload yet wastes a high amount of data centers energy.
FPGA virtualization attempted to resolve this issue by splitting the FPGA fabric into multiple chunks and implementing applications in the so-called virtual FPGAs.
Yazdanshenas \emph{et al.} have quantified the cost of FPGA virtualization in \cite{yazdanshenas2017quantifying}, revealing up to 46\% performance degradation with 2.6$\times$ increase in wire length of the \emph{shell}, i.e., the static region responsible to connect the virtual FPGAs to external resources such as PCI and DDR.
This hinders the routability of the shell as the number of virtual FPGAs increase.
These overheads excluded the area overhead of the shell itself, which occupies up to 44\% of FPGA area.
FPGA virtualization is also not practical for large data center applications such as deep neural networks that occupy a whole or multiple devices \cite{ovtcharov2015accelerating}.

Another foray for FPGA power optimization includes approaches that exploit dynamic frequency and/or voltage scaling.
The main goal of these studies is to utilize the available timing headroom conservatively considered for worst-case temperature, aging, variation, etc. and scale the frequency for performance boosting, or voltage reduction without performance degradation, though a few of them consider workload.
Chow \emph{et al.} \cite{chow2005dynamic} propose a dynamic voltage scaling scheme that exploits a ring-oscillator based logic delay measurement circuit to mimic the timing behavior of application critical path and adjust the voltage accordingly. However, the inaccuracy of path monitor circuitries in FPGAs and even ASICs has been well elaborated \cite{levine2014dynamic, zhao2016universal, amrouch2016reliability, ahmadi2018dynamic}.
Levine \emph{et al.} employ timing error detectors inserted as capture registers with a phase-shifted clock at the end of critical paths to find out the timing slack of FPGA-mapped designs through a gradual reduction of voltage \cite{levine2014dynamic}.
Their approach adds extra area and power overhead, cannot be implemented in paths heading to hard blocks such as memories, and assumes the corresponding paths will be exercised at runtime.
Zhao \emph{et al.} propose an elaborated two-step approach by extracting the critical paths of the design using the static timing analysis tool and sequentially mapping into the FPGA \cite{zhao2016universal}.
Thence, they vary the FPGA core voltage to obtain the voltage-delay ($V_{core}-D$) relation of the paths for online adjustment during the operation time.
It requires analyzing a huge number of paths, especially originally non-critical paths might become critical when the voltage changes.
Salami \emph{et al.} evaluate the impact of block RAM (BRAM) voltage ($V_{bram}$) scaling on the power and accuracy of a neural network application \cite{salami2018comprehensive}. They observed that $V_{bram}$ can be reduced by 39\% of the nominal value, which saves the BRAM dynamic power by one order of magnitude, with a negligible error at the output.
Their approach is intuitive and does not examine timing violation, i.e., it is not known if the timing will not be eventually violated in a particular voltage level.
Similarly, Khaleghi \emph{et al.} leverage the thermal margin of FPGAs for frequency boosting though they integrate it in the conventional flow of FPGA using the pre-characterizition of resources \cite{khaleghi2019thermal}.
Eventually, Jones \emph{et al.} propose a workload-aware frequency scaling approach that temporarily allows over-clocking of applications when the temperature is safe enough, i.e., the workload is not bursty \cite{jones2007dynamically}. They assume the design has inherently sufficient slack to tolerate the frequency boosting without overscaling the voltage.

As mentioned earlier, the primary goal of the latter studies is to leverage the pessimistic timing headrooms for efficiency, while they struggle in guaranteeing timing safety.
More importantly, the utmost effort of previous works is to satisfy the timing of critical or near-critical paths under either (and mainly) $V_{core}$ scaling, or $V_{bram}$ scaling.
Nevertheless, unlike single voltage scaling where there is only one minimal voltage level for a target frequency, for simultaneous scaling of $V_{core}$ and $V_{bram}$, numerous `$V_{core}$, $V_{bram}$' pairs will minimally yield the target frequency while only one pair of this solution space has the minimum power dissipation.
Therefore, accurate timing \emph{and power} analysis under multiple voltage scaling is inevitable.

\begin{figure}[t]
  \centering
  \includegraphics[width=0.5\textwidth]{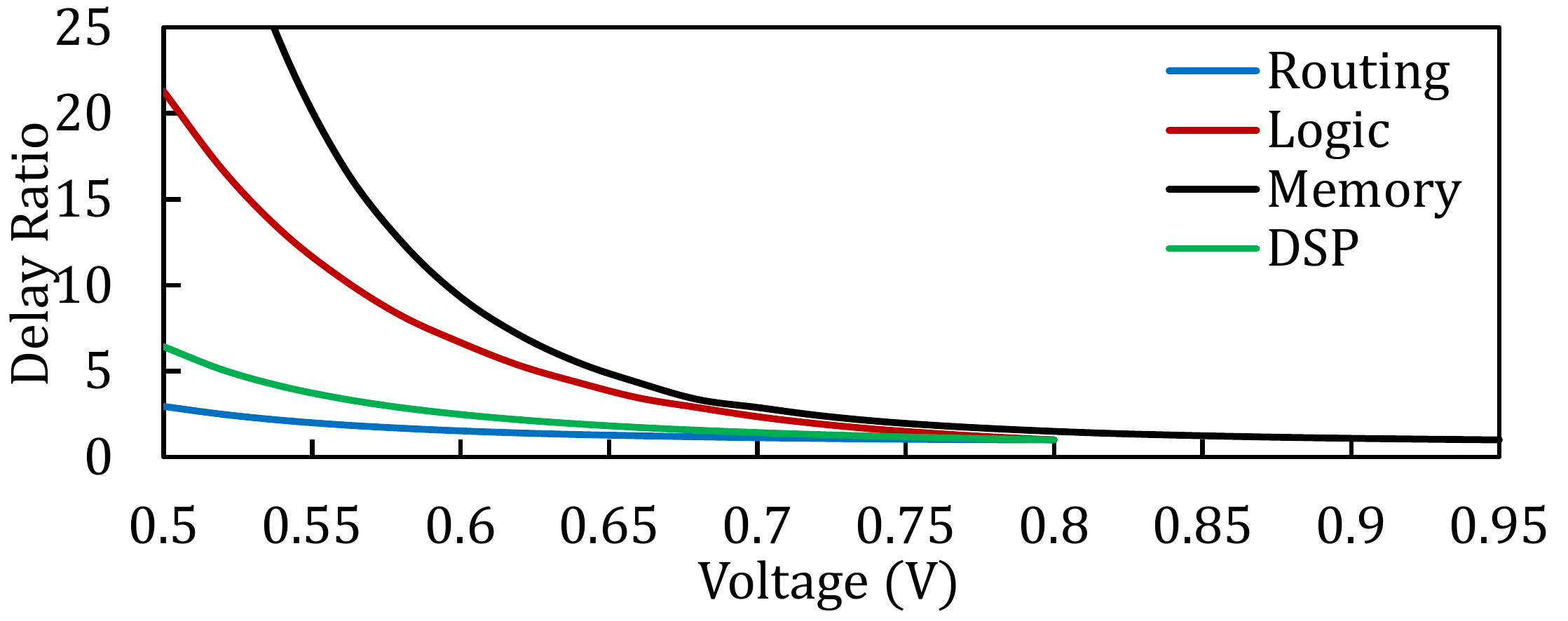}
  \caption{ Delay of FPGA resources versus voltage $\downarrow$}  \label{fig:D-V}
\end{figure}

\begin{figure}[t]
  \centering
  \includegraphics[width=0.5\textwidth]{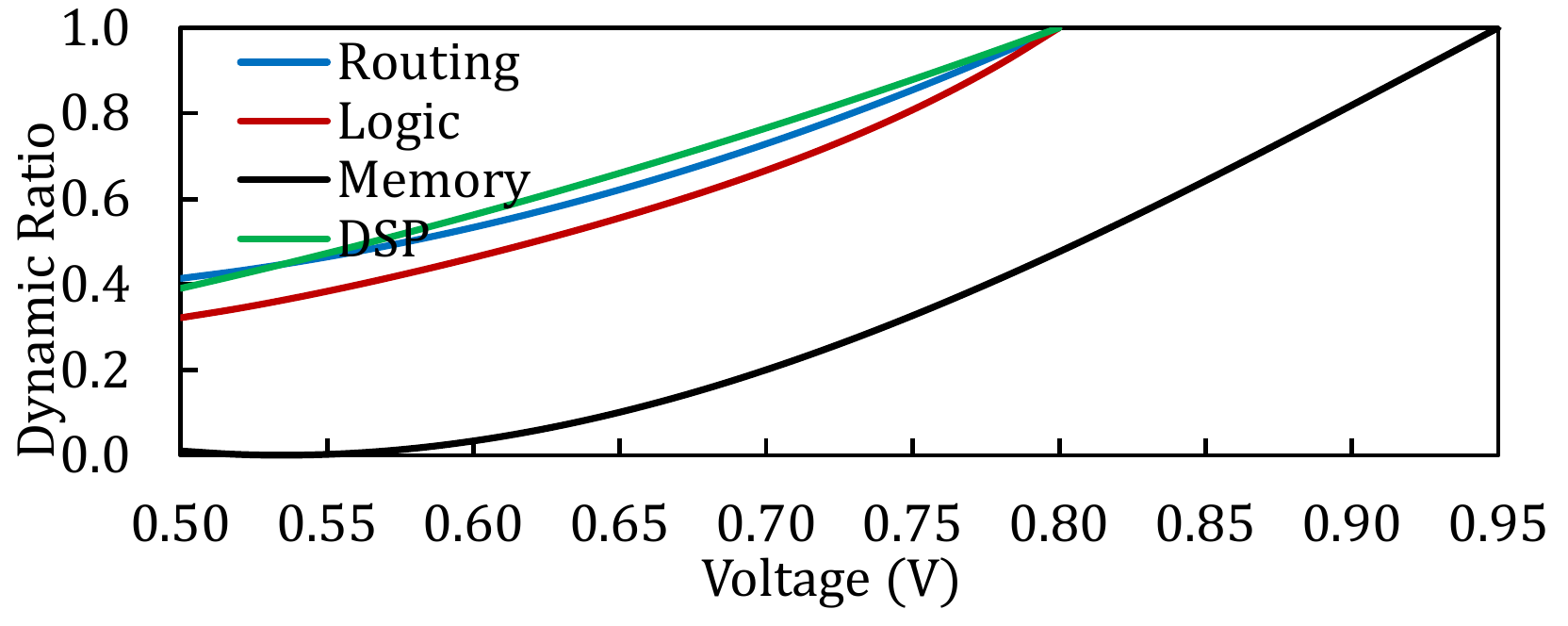}
  \caption{ Dynamic power of FPGA resources versus voltage.} \label{fig:Dynamic-V}
\end{figure}

\begin{figure}[t]
  \centering
  \includegraphics[width=0.5\textwidth]{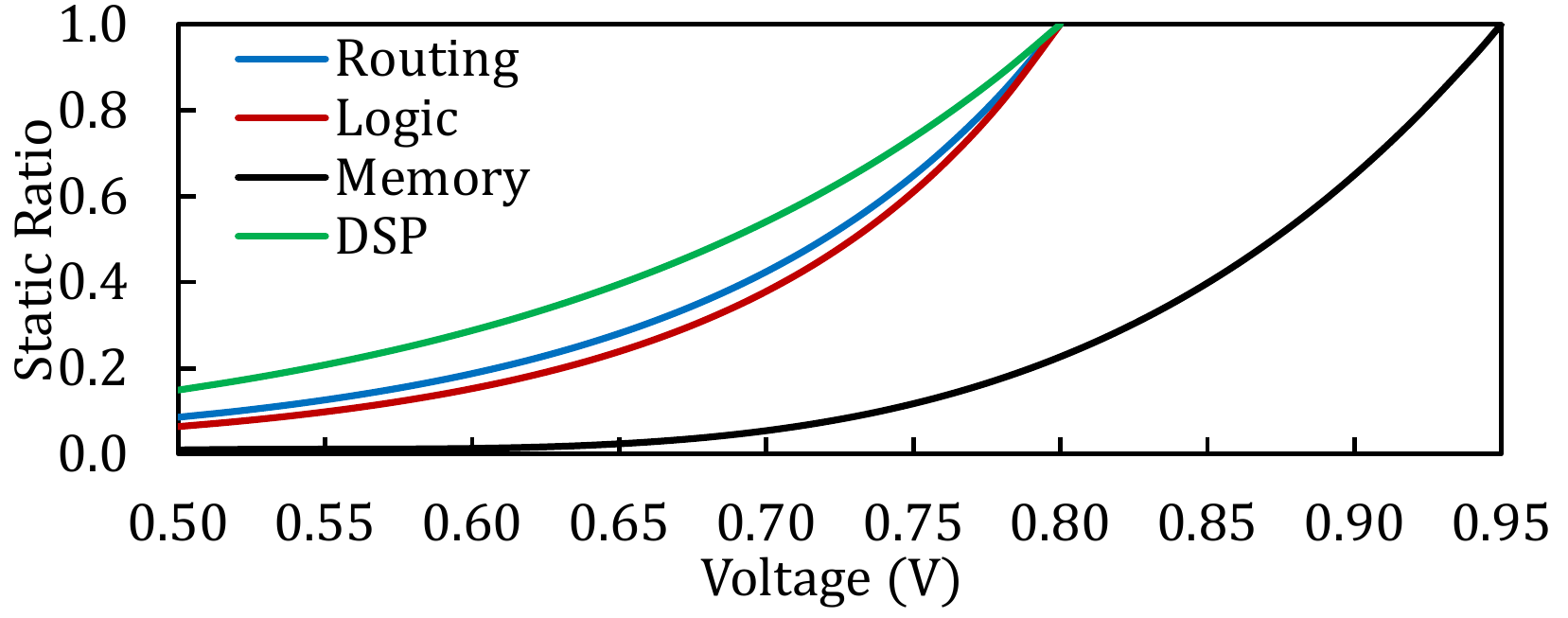}
  \caption{ Static power of FPGA resources versus voltage.} \label{fig:static-V}
\end{figure}





\section{Motivational Analysis} \label{sec:motivation}
In this section, we use a simplified example to justify the necessity of the proposed scheme and how it surpasses the conventional approaches in power efficiency.
Figures \ref{fig:D-V}, \ref{fig:Dynamic-V}, \ref{fig:static-V} shows the relation of delay and power consumption of FPGA resources when voltage scales down.
Experimental results will be elaborated in Section \ref{sec:result}, but concisely, routing and logic delay and power indicate the average delay and power of individual routing resources (e.g., switch boxes and connection block multiplexers) and logic resources (e.g., LUTs).
Memory stands for the on-chip BRAMs, and DSP is the digital signal processing hard macro block.
Except memory blocks, the other resources share the same $V_{core}$ power rail.
Since FPGA memories incorporate high-threshold process technology, they utilize a $V_{bram}$ voltage that is initially higher than nominal core voltage $V_{core}$ to enhance the performance \cite{yazdanshenas2017don}.
We assumed a nominal memory and core voltage of 0.95V and 0.8V, respectively \cite{yazdanshenas2017don}.

The different sensitivity of resources' delay and power with respect to voltage scaling implies cautious considerations when scaling the voltage. For instance, by comparing Figure \ref{fig:D-V} and Figure \ref{fig:static-V}, we can understand that reducing the memory voltage from 0.95V down to 0.80V has a relatively small effect on its delay, while its static power decreases by more than 75\%.
Then we see a spike in memory delay with trivial improvement of its power, meaning that it is not beneficial to scale $V_{bram}$  anymore.
Similarly, routing resources show good delay tolerance versus voltage scaling. It is mainly because of their simple two-level pass-transistor based structure with boosted configuration SRAM voltage that alleviates the drop of drain voltages \cite{chiasson2013coffe}.
Notice that we assume a separate power rail for configuration SRAM cells and do not change their voltages as they are made up of thick high-threshold transistors that have already throttled their leakage current by two orders of magnitude though have a crucial impact on FPGA performance. Nor we do scale the auxiliary voltage of I/O rails to facile standard interfacing.
While low sensitivity of routing resources against voltage implied $V_{core}$ is a prosperous candidate in interconnection-bound designs, the large increase of logic delay with voltage scaling hinders $V_{core}$ scaling when the critical path consists of mostly LUTs.
In the following we show how varying parameters of workload, critical path(s), and application affect optimum `$V_{core}$, $V_{bram}$' point and energy saving.

Let us consider the critical path delay of an arbitrary application as Equation \eqref{eq:cp}.
\begin{equation}\label{eq:cp}
    d_{cp} = d_{l0} \cdot \mathcal{D}_{l}(V_{core}) + d_{m0} \cdot \mathcal{D}_{m}(V_{bram})
\end{equation}
Where $d_{l0}$ stands for the initial delay of the logic and routing part of the critical path, and $\mathcal{D}_{l}(V_{core})$ denotes the voltage scaling factor, i.e., information of Figure \ref{fig:D-V}.
Analogously, $d_{m0}$ and $\mathcal{D}_{m}(V_{bram})$ are the memory counterparts.
The original delay of the application is $d_{l0} + d_{m0}$, which can be stretched by $(d_{l0} + d_{m0}) \times S_{w}$ where $S_{w} \geq 1$
indicates the workload factor, meaning that in an 80\% workload, the delay of all nodes can be increased up to $S_{w} = \frac{1}{0.8} = 1.25\times$.
Defining $\alpha = \frac{d_{m0}}{d_{l0}}$ as the relative delay of memory block(s) in the critical path to logic/routing resources, the applications need to meet the following:
\begin{equation}\label{eq:cp2}
    d_{cp} \propto \mathcal{D}_{l}(V_{core}) + \alpha \cdot \mathcal{D}_{m}(V_{bram}) \leq (1 + \alpha) \cdot S_{w}
\end{equation}

We can derive a similar model for power consumption as a function of $V_{core}$ and $V_{bram}$ shown by Equation \eqref{eq:power}.
\begin{equation}\label{eq:power}
    p_{cir} \propto \mathcal{P}_{l}(V_{core}, d_{cp}) + \beta \cdot \mathcal{P}_{m}(V_{bram}, d_{cp})
\end{equation}
where $\mathcal{P}_{l}(V_{core}, d_{cp})$ is for the total power drawn from the core rail by logic, routing, and DSP resources as a function of voltage $V_{core}$ and frequency (delay) $d_{cp}$, and $\beta$ is an application-dependent factor to determine the contribution of BRAM power.
In the following, we initially assume $\alpha = 0.2$ (i.e., BRAM contributes to $\frac{0.2}{1+0.2}$ of critical path delay \cite{chiasson2013coffe}) and $\beta = 0.4$ (i.e., BRAM power initially is $\sim 25\%$ of device total power \cite{salami2018comprehensive}).

\begin{figure}[t]
  \centering
  \includegraphics[width=0.5\textwidth]{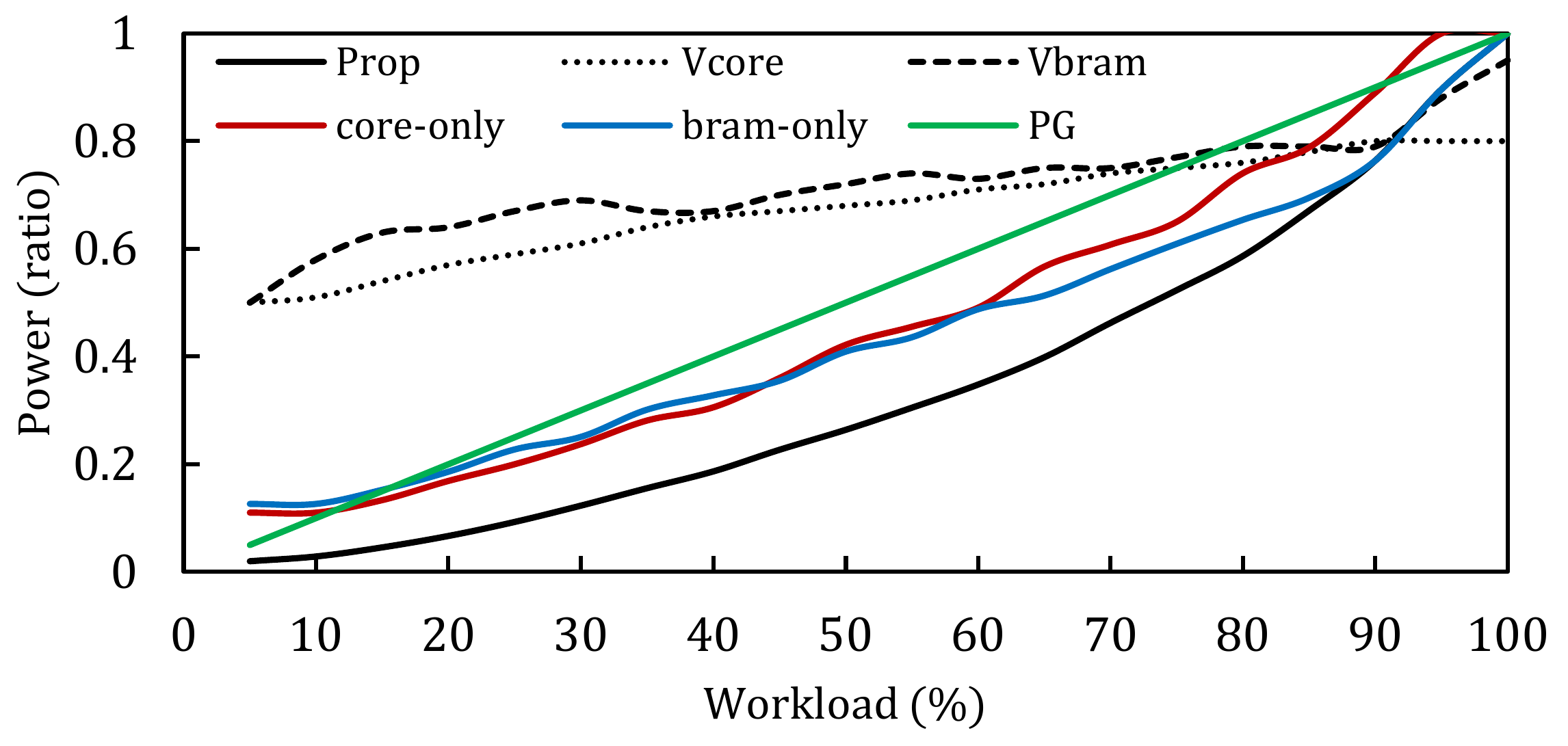}
  \caption{Comparing DVFS techniques in different workloads $\downarrow$} \label{fig:P-W}
\end{figure}
\begin{figure}[t]
  \centering
  \includegraphics[width=0.5\textwidth]{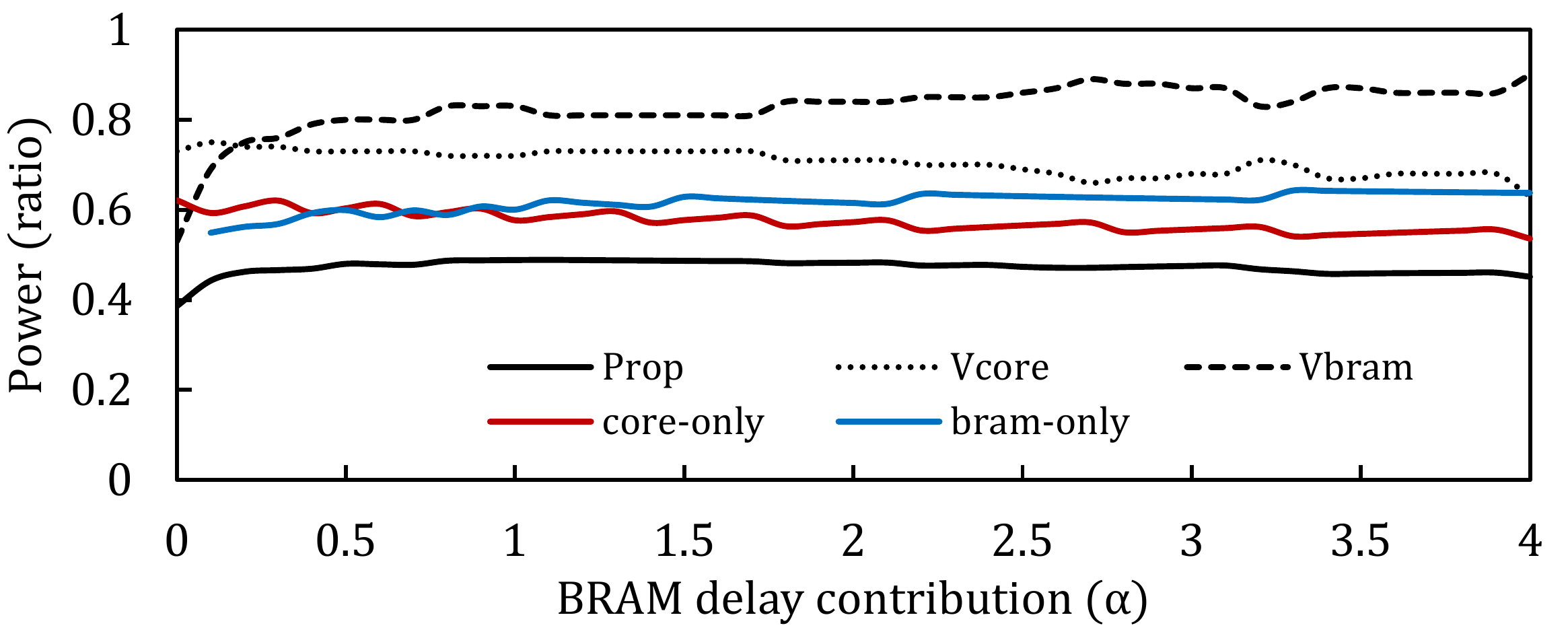}
  \caption{Comparing DVFS techniques in different critical paths.} \label{fig:P-alpha}
\end{figure}
\begin{figure}[t]
  \centering
  \includegraphics[width=0.5\textwidth]{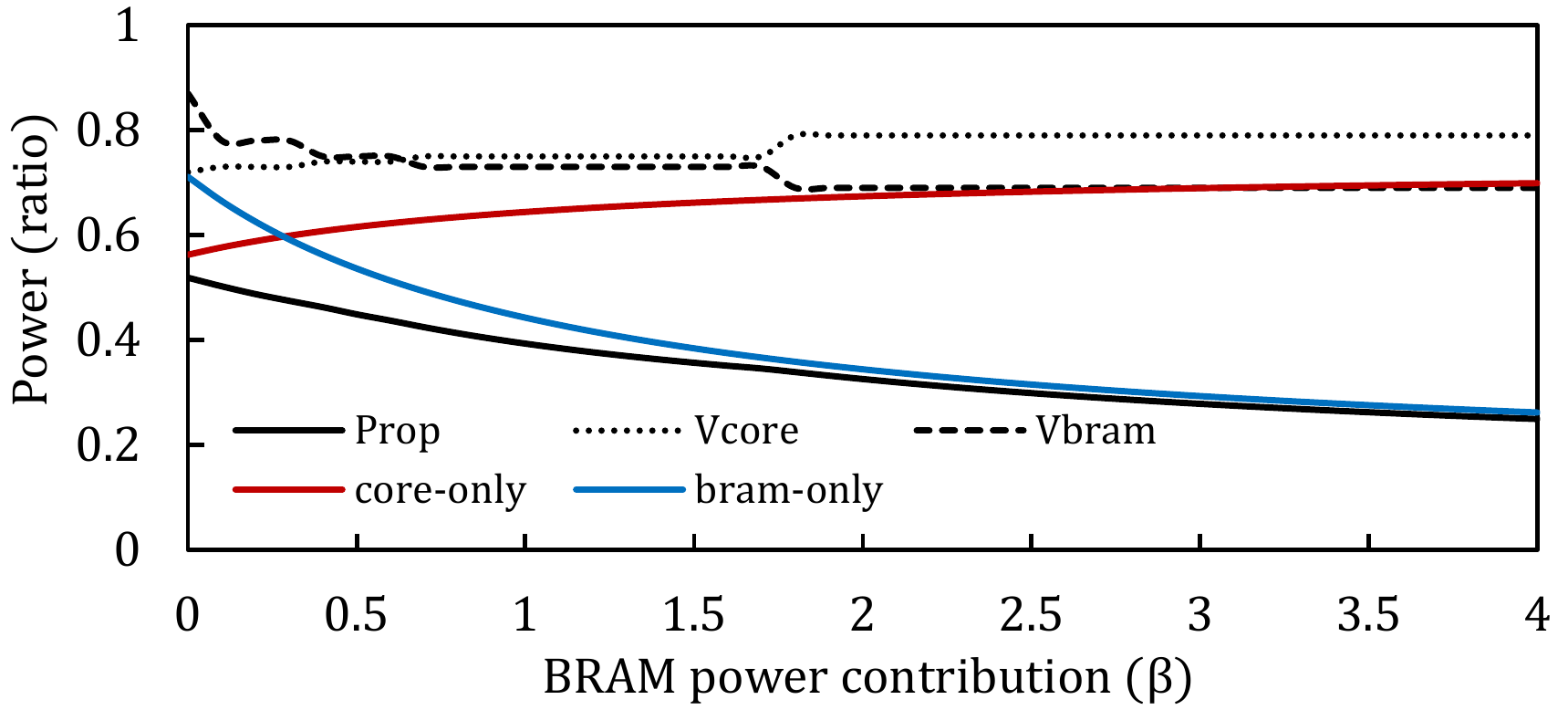}
  \caption{Comparing DVFS techniques in different BRAM power rates.} \label{fig:P-beta}
\end{figure}





Figures \ref{fig:P-W}, \ref{fig:P-alpha}, \ref{fig:P-beta}  demonstrates the efficiency of different voltage scaling schemes under varying workloads, applications' critical paths (`$\alpha$'s), and applications' power characteristics (i.e., $\beta$, the ratio of memory to chip power).
\emph{Prop} means the proposed approach that simultaneously determines $V_{core}$ and $V_{bram}$, core-only is the technique that only scales $V_{core}$ \cite{zhao2016universal, levine2014dynamic}, and bram-only is similar to \cite{salami2018comprehensive}.
Dashed lines of \emph{Vcore} and \emph{Vbram} in the figures show the \emph{magnitude} of the $V_{core}$ and $V_{bram}$ in the proposed approach, \emph{Prop} (for the sake of clarity, we do not show voltages of the other methods).
According to Figure \ref{fig:P-W}, in high workloads ($> 90\%$, or $S_{w} < 1.1$), our proposed approach mostly reduces the $V_{bram}$ voltage because slight reduction of the memory power in high voltages significantly improves the power efficiency, especially because the contribution of memory delay in the critical path is small ($\alpha = 0.2$), leaving room for $V_{bram}$ scaling. For the same reason, core-only scheme has small gains there.
The Figure also reveals the  sophisticated relation of the minimum voltage points and the size of workload; each workload level requires re-estimation of `$V_{core}, V_{bram}$'.
In all cases, the proposed approach yields the lowest power consumption.
It is noteworthy that the conventional power-gating approach (denoted by \emph{PG} in Figure \ref{fig:P-W}) scales the number of \emph{computing nodes} linearly with workload, though, the other approaches scale both frequency and voltage, leading to twofold power saving. 
In very low workloads, power-gating works better than the other two approaches because the crash voltage ($\sim 0.50V$) prevents further power reduction.

Similar insights can also be grasped from Figure \ref{fig:P-alpha} and \ref{fig:P-beta}. A constant workload of 50\% is assumed here while $\alpha$ and $\beta$ parameters change.
When the contribution of BRAM delay in total reduces, the proposed approach tends to scale the $V_{bram}$. For $\alpha = 0$ highest power saving is achieved as the proposed method can scale the voltage to the minimum possible, i.e., the crash voltage.
Analogously in Figure \ref{fig:P-beta}, the effectiveness of the core-only (bram-only) method degrades (improves) when BRAM contributes to a significant ratio of total power, while our proposed method can adjust both voltages cautiously to provide minimum power consumption.
It is worth to note that the efficiency of the proposed method increases in high BRAM powers because in these scenarios a minor reduction of BRAM power saves huge power with a small increase of delay (compare Figure\ref{fig:P-W} and  \ref{fig:P-beta}).

\section{Proposed Method}\label{sec:method}

\begin{figure}[t]
  \centering
  \includegraphics[width=0.5\textwidth]{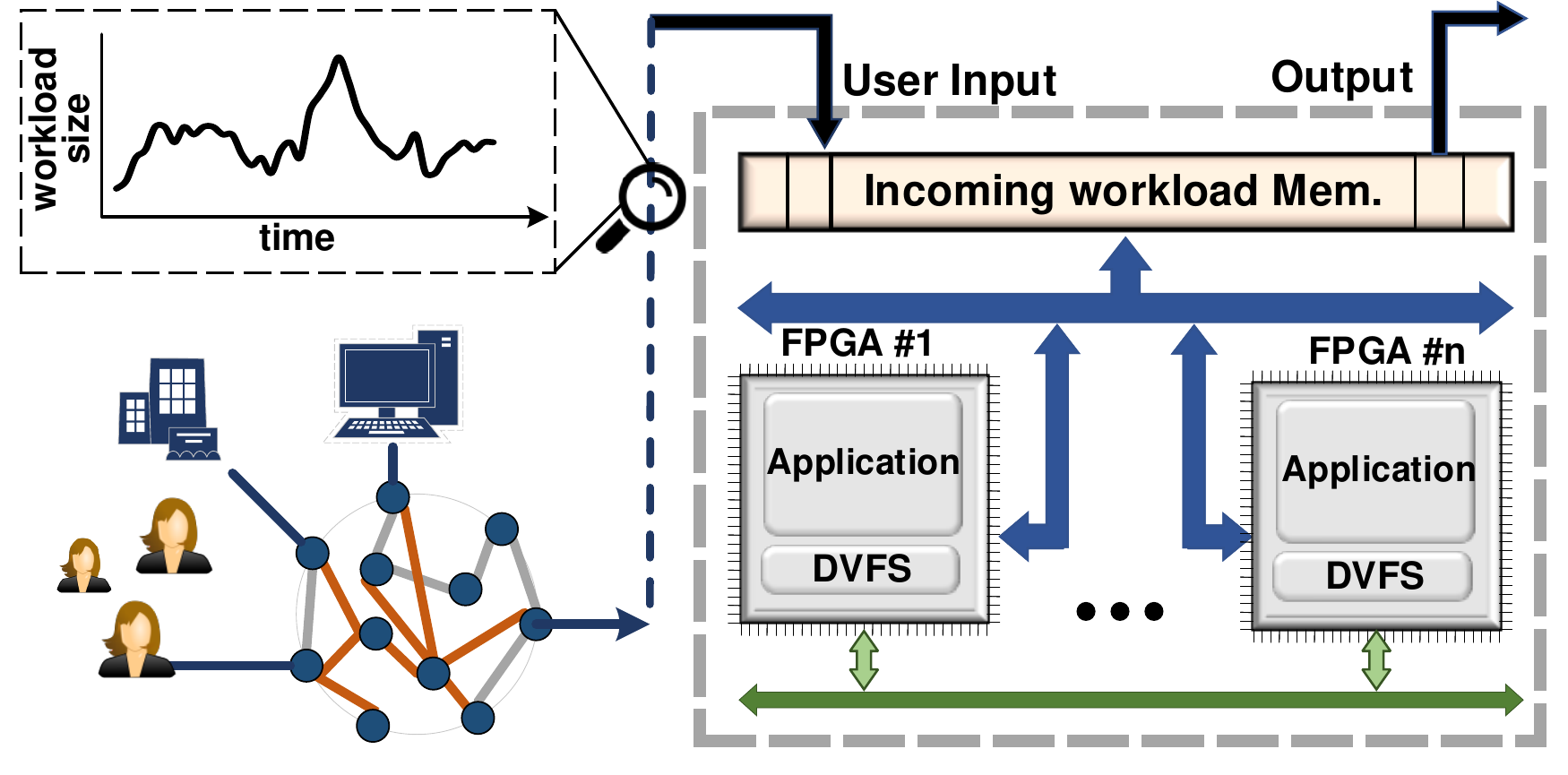}
  \caption{Overview of an FPGA-based datacenter platform.}  \label{fig:platform}
  \vspace{-0.6cm}
\end{figure}

In practice, the generated data from different users are processed in a centralized FPGA platform located in datacenters. The computing resources of the data centers are rarely completely idle and sporadically operate near their maximum capacity. In fact, most of the time the incoming workload is between 10\% to 50\% of the maximum nominal workload. Multiple FPGA instances are designed to deliver the maximum nominal workload when running on the nominal frequency to provide the users' desired quality of service. 
However, since the incoming FPGA workloads are often lower than the maximum nominal workload, FPGA become underutilized.
By scaling the operating frequency proportional to the incoming workload, the power dissipation will be reduced without violating the desired throughput. It is noteworthy that if an application has specific latency restrictions, it should be considered in the voltage and frequency scaling.
The maximum operating frequency of the FPGA can be set depending on the delay of the critical path such that it guarantees the reliability and the correctness of the computation.
By underscaling the frequency, i.e., stretching the clock period, delay of the critical path becomes less than the clock toggle rate.
This extra timing room can be leveraged to underscale the voltage to minimize the energy consumption untill the critical path delay again reaches the clock delay.

Figure \ref{fig:platform} abstracts an FPGA cloud platform consisting of $n$ FPGA instances where all of them are processing the input data gathered from one or different users. FPGA instances are provided with the ability to modify their operating frequency and voltage.
In the following we explain the workload prediction, dynamic frequency scaling and dynamic voltage scaling implementations.

\subsection{Workload Prediction}
 We divide the FPGA execution time to steps with the length of $\tau$, where the energy is minimized separately for each time step. 
At the $i^{th}$ time step ($\tau_{i-1}$), our approach predicts the size of the workload for the $i+1$ time step.
Accordingly, we set the working frequency of the platform such that it can complete the the predicted workload for the $\tau_{i}$ time step.

To provide the desired QoS as well as minimizing the FPGA idle time, the size of the incoming workload needs to be predicted at each time step.
The operating voltage and frequency of the platform is set based on the predicted workload.
Generally, to predict and allocate resources for dynamic workloads, two different approaches have been established: reactive, and proactive.
In reactive approach, resources are allocated to the workload based on a predefined thresholds \cite{bonvin2011autonomic, zhu2010resource}, while in proactive approach, the future size of the workload is predicted and resources are allocated based on this prediction \cite{islam2012empirical, calheiros2015workload, gong2010press}. 

In this work, we use a light-weight online workload prediction method similar to the one proposed in \cite{gong2010press} which is able to extract short-term features.
In the cases the service provider knows the periodic signatures of the incoming workload, the predictor can be loaded with this information.
Workloads with repeating patterns are divided into time intervals which are repeated with the period.
The average of the intervals represents a bias for the short-term prediction.
For applications without repeating patterns, we use a discrete-time Markov chain with a finite number of states to represents the short-term characteristics of the incoming workload.

\begin{figure}[t]
  \centering
  \includegraphics[width=0.4\textwidth]{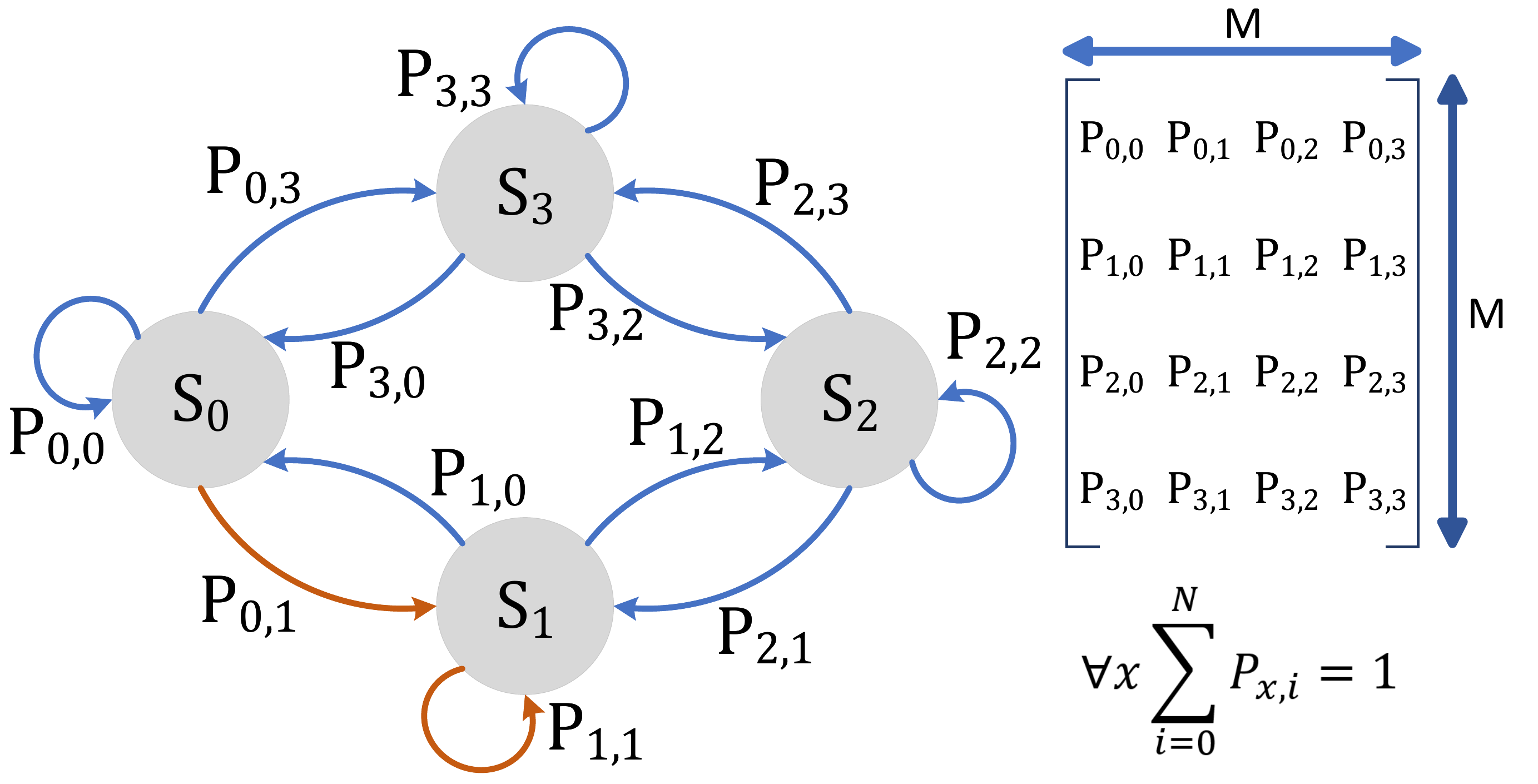}
  \caption{Example of Markov chain for workload prediction.}  \label{fig:markov}
  \vspace{-0.7cm}
\end{figure}

The size of the workload is discretized into $M$ bins, each represented by a state in the Markov chain; all the states are connected through a directed edge. $P_{i,j}$ shows the transition probability from state $i$ to state $j$.
Therefore, there are $M \times M$ edges between states where each edge has a probability learned during the training steps to predict the size of the incoming workload. Figure \ref{fig:markov} represents a Markov chain model with 4 states, $\{S_0, S_1, S_2, S_3\}$, in which a directed edge with label $P_{i,j}$ shows the transition from $S_i$ to $S_j$ which happens with the probability of $P_{i,j}$. Considering the The total probability of the outgoing edges of state $S_i$ has to be 1 as probability of selecting the next state is one.

Starting from $S_0$ with probability of $P_{0,i}$ the next state will be $S_i$.
In the next time step, the third state will be again $S_1$ with $P_{1,1}$ probability.
If a pre-trained model of the workload is available, it can be  loaded on FPGA, otherwise, the model needs to be trained during the runtime.
During system initialization, the platform runs with the maximum frequency and works with the nominal frequency for the first $I$ time steps. In the training phase, the Markov model learns the patterns of the incoming workload and the probability of transitions between states are set during this phase.

After $I$ time steps, the Markov model predicts the incoming input of the next time step and the frequency of the platform is selected accordingly, with a $t\%$ throughput margin to offset the likelihood of workload under-estimation as well as to preclude consecutive mispredictions.
Mispredictions can be either under-estimations or over-estimations.
In case of over-estimation, QoS is meet, however, some power is wasted as the frequency (and voltage) is set to a unnecessarily higher value.
In case of workload under-estimation the desired QoS may be violated.
The work in \cite{gong2010press} tackles most of the underestimations by $t=5\%$ margin.

\subsection{Frequency Scaling Flow}
To achieve high energy efficiency, the operating FPGA frequency needs to be adjusted according to the size of the incoming workload.
To scale the frequency of FPGAs, Intel (Altera) FPGAs enable  Phase-Locked Loop (PLL) hard-macros (Xilinx also provide a similar feature).
Each PLL generates up to 10 output clock signals from a reference clock. Each clock signal can have an independent frequency and phase as compared to the reference clock. 
PLLs support runtime reconfiguration through a Reconfiguration Port (RP).
The reconfiguration process is capable of updating most of the PLL specifications, including clock frequency parameters sets (e.g. frequency and phase). To update the PLL parameters, a state machine controls the RP signals to all the FPGA PLL modules.

PLL module has a \textit{Lock} signal that represents when the output clock signal is stable.
The lock signal activates whenever there is a change in PLL inputs or parameters. After stabling the PLL inputs and the output clock signal, the lock signal is asserted again.
The lock signal is de-asserted during the PLL reprogramming and will be issued again in, at most, $100 \mu Sec$.
Each of the FPGA instances in the proposed DFS module has its own PLL modules to generate the clock signal from the reference clock provided in the FPGA board.
For simplicity of explanations, we assume the design works with one clock frequency, however, our design supports multiple clock signals with the same procedure. 
Each PLL generates one clock output, \textit{CLK0}. At the start-up, the PLL is initialized to generate the output clock equal to the reference clock.
When the platform modifies the clock frequency, at $\tau_i$ based on the predicted workload for $\tau_{i+1}$, the PLL is reconfigured to generate the output clock that meets the QoS for $\tau_{i+1}$.

\subsection{Voltage Scaling Flow}
To implement the dynamic voltage scaling for both $V_{core}$ and $V_{bram}$, Texas Instruments (TI) PMBUS USB Adapter can be used \cite{TI} for different FPGA vendors.
TI adapter provides a C-based Application Programming Interface (API), which eases adjusting the board voltage rails and reading the chip currents to measure the power consumption through Power Management Bus (PMBUS) standard.
To scale the FPGA voltage rails, the required PMBUS commands are sent to the adapter to set the $V_{bram}$ and $V_{core}$ to certain values.
This adopter is used as a proof of concept, while in industry fast DC-DC converters are used to change the voltage rails.
The work in \cite{jain20140} has shown a latency of 3-5 nSec, and is able to generate voltages between 0.45V to 1V with 25mV resolution.
As these converters are faster than the FPGAs clock frequency, we neglect the performance overhead of the DVS module in the rest of the paper.

\begin{figure*}[t]
  \centering
  \vspace{-0.5cm}
  \includegraphics[width=1\textwidth]{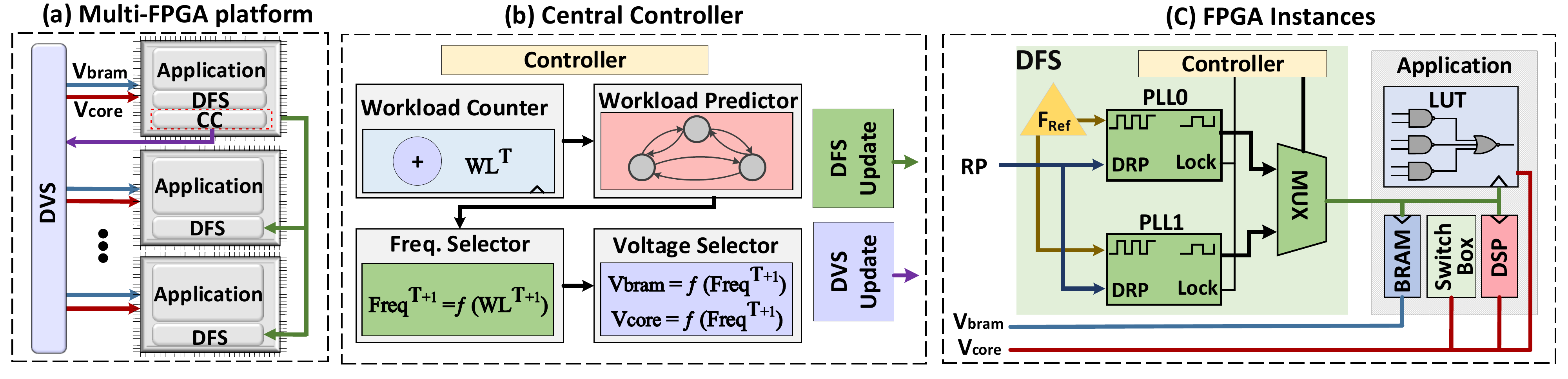}
  \vspace{-0.4cm}
  \caption{(a) the architecture of the proposed energy-efficient multi-FPGA platform. The details of the (b) central controller, and (c) the FPGA instances.} \label{fig:arch}
  \vspace{-0.5cm}
\end{figure*}

\section{Proposed Architecture}
Figure \ref{fig:arch}(a) demonstrates the architecture of the proposed energy efficient multi-FPGA platform. 
Our platform consists of $n$ FPGAs where one of them is a central FPGA. The central FPGA has Central Controller (CC) and DFS blocks and is responsible to control the frequency and voltage of all other FPGAs. 
Figure~\ref{fig:arch}(b) shows the details of the CC managing the voltage/frequency of all FPGA instances. The CC predicts the workload size and accordingly scales the voltage and frequency of all other FPGAs.
A \textit{Workload Counter} computes the number of incoming inputs in a central FPGA, assuming  all other FPGAs have the similar input rate. 
The \textit{Workload Predictor} module compares the counter value with the predicted workload at the previous time step. Based on the current state, the workload predictor estimates the workload size in the next time step. 
Next, \textit{Freq. Selector} module determines the frequency of all FPGA instances depending on the workload size.
Finally, the \textit{Voltage Selector} module sets the working voltages of different blocks based on the clock frequency, design timing characteristics (e.g., critical paths), and FPGA resources characteristics.
This voltage selection happens for logic elements, switch boxes, and DSP cores ($V_{core}$); as well as the operating voltage of BRAM cells ($V_{bram}$). The obtained voltages not only guarantee timing (which has a large solution space), but also minimizes the power as discussed in Section \ref{sec:motivation}.
The optimal operating voltage(s) of each frequency is calculated during the design synthesis stage and are stored in the memory, where the DVS module is programmed to fetch the voltage levels of FPGAs instances.

\textbf{Misprediction Detection: }
In CC, the misprediction happens when the workload bin for time step $i^{th}$ is not equal to the bin achieved by the workload counter. 
To detect mispredictions, the value of $t\%$ should be greater than $1/m$, where $m$ is the number of bins.
Therefore, the system discriminates each bin with the higher level bin. For example, if the size of the incoming workload is predicted to be in bin $i^{th}$ while it actually belongs to $i+1^{th}$ bin, the system is able to process the workload with the size of $i+1^{th}$ bin. 
After each misprediction, the state of the Markov model is updated to the correct state.
If the number of mispredictions exceeded a threshold, the probabilities of the corresponding edges are updated.

\textbf{PLL Overhead: }
The CC issues the required signals to reprogram the PLL blocks in each FPGA.
To reprogram the PLL modules, the DVF reprogramming FSM issues the RP signal serially. After reprogramming the PLL module, the generated clock output is unreliable until the lock signal is issued, which takes no longer than 100 $\mu Sec$.
In the cases the framework changes the frequency and voltage very frequently, the overhead of stalling the FPGA instances for the stable output clock signal limits the performance and energy improvement.
Therefore, we use two PLL modules to eliminate the overhead of frequency adjustion.
In this platform, as shown in Figure \ref{fig:arch}(c), the outputs of two PLL modules pass through a multiplexer, one of them is generating the current clock frequency, while the other is being programmed to generate the clock for the next time step.
Thus, in the next clock, the platform will not be halted waiting for a stable clock frequency.

In case of having one PLL, each time step with duration $\tau$ requires $t_{lock}$ extra time for generating a stable clock signal. 
Therefore, using one PLL has $t_{lock}$ set up overhead. 
Since $t_{lock} \ll \tau$, we assume the PLL overhead, $t_{lock}$, does not affect the frequency selection. 
The energy overhead of using one PLL is:
\begin{equation}
\underbrace{P_{Design} \times t_{lock}}_{Design \ energy \ during \ t_{lock}} + \underbrace{P_{PLL} \times (\tau + t_{lock})}_{PLL \ energy}
\end{equation}
In case of using two PLLs, there is no performance overhead. The energy overhead would be equal to power consumption of two PLLs multiplied by $\tau$. The performance overhead is negligible since $t_{lock} < 100\mu Sec \ll \tau$.
Therefore, it is more efficient to use two PLLs when the following condition is hold:
\begin{equation}
P_{design} \times t_{lock} + P_{PLL} \times (\tau + t_{lock}) > 2 \times P_{PLL} \times \tau
\end{equation}
Since $t_{lock} \ll \tau$, we should have $P_{design} \times t_{Lock} > P_{PLL} \times \tau$.
Our evaluation shows that this condition can be always satisfied over all our experiments. 
In practice, the fully utilized FPGA power consumption is around 20W while the PLL consumes about 0.1W, and $t_{lock} \simeq 10 \mu Sec$.
Therefore, when $\tau > 2 mSec$, the overhead of using two PLL becomes less than using one PLL. In practice, $\tau$ is at least in order of seconds or minutes; thus it is always more beneficial to use two PLLs. 

\section{Experimental Results} \label{sec:result}
\subsection{General Setup}
We evaluated the efficiency of the proposed method by implementing several state-of-the-art neural network acceleration frameworks on a commercial FPGA architecture.
To generate and characterize the SPICE netlist of FPGA resources from delay and power perspectives, we used the latest version of COFFE \cite{yazdanshenas2019coffe} with 22nm predictive technology model (PTM) \cite{ptm} and an architectural description file similar to Stratix IV devices due to their well-provided architectural details \cite{luu2014vtr}.
COFFE does not model DSPs, so we hand-crafted a Verilog HDL of Stratix IV DSPs \cite{stratixiv} and characterized with Synopsys Design Compiler using NanGate 45nm Open-Cell Library \cite{nangate2008california} tailored for libraries with different voltages by the means of Synopsys SiliconSmart.
Eventually we scaled the 45nm DSP characterization to 22nm following the scaling factors of a subset of combinational and sequential cells obtained through SPICE simulations.

We synthesized the benchmarks using Intel (Altera) Quartus II software targeting Stratix IV devices and converted the resulted VQM (Verilog Quartus Mapping) file format to Berkeley Logic Interchange Format (BLIF) format, recognizable by our placement and routing VTR (Verilog-to-Routing) toolset \cite{luu2014vtr}.
VTR gets a synthesized design in BLIF format along with the architectural description of the device (e.g., number of LUTs per slice, routing network information such as wire length, delays, etc.) and maps (i.e., performs place and routing) on the smallest possible FPGA device and simultaneously tries to minimize the delay.
The only amendment we made in the device architecture was to increase the capacity of I/O pads from 2 to 4 as our benchmarks are heavily I/O bound.
Our benchmarks include Tabla \cite{mahajan2016tabla}, DnnWeaver \cite{sharma2016high}, DianNao \cite{chen2014diannao}, Stripes \cite{judd2016stripes}, and Proteus \cite{judd2016proteus} which are general neural network acceleration frameworks capable of optimizing various objective functions through gradient descent by supporting huge parallelism. The last two networks provide serial and variable-precision acceleration for energy efficiency.
Table \ref{tab:synthesis} summarizes the resource usage and post place and route frequencies of the synthesized benchmarks.
LAB stands for Logic Array Block and includes 10 6-input LUTs.
M9K and M144K show the number of 9Kb and 144Kb memories.

\begin{table}[t]
\caption{Post place and route resource utilization and timing of the benchmarks.}
  \vspace{-0.2cm}
\resizebox{0.5\textwidth}{!}{
\begin{tabular}{llllll}
\hline
\textbf{Parameter} & \textbf{Tabla} & \textbf{DnnWeaver} & \textbf{DianNao} & \textbf{Stripes} & \textbf{Proteus} \\ \hline
LAB    & 127   & 730       & 3430    & 12343   & 2702    \\ 
DSP    & 0     & 1         & 112     & 16      & 144     \\ 
M9K    & 47    & 166       & 30      & 15      & 15      \\ 
M144K  & 1     & 13        & 2       & 1       & 1       \\ 
I/O    & 567   & 1655      & 4659    & 8797    & 5033    \\
Freq. (MHz)   & 113   & 99      & 83    & 40    & 70    \\ \hline
\end{tabular} \label{tab:synthesis}
}
\end{table}

\begin{figure}[t]
  \centering
  \includegraphics[width=0.5\textwidth]{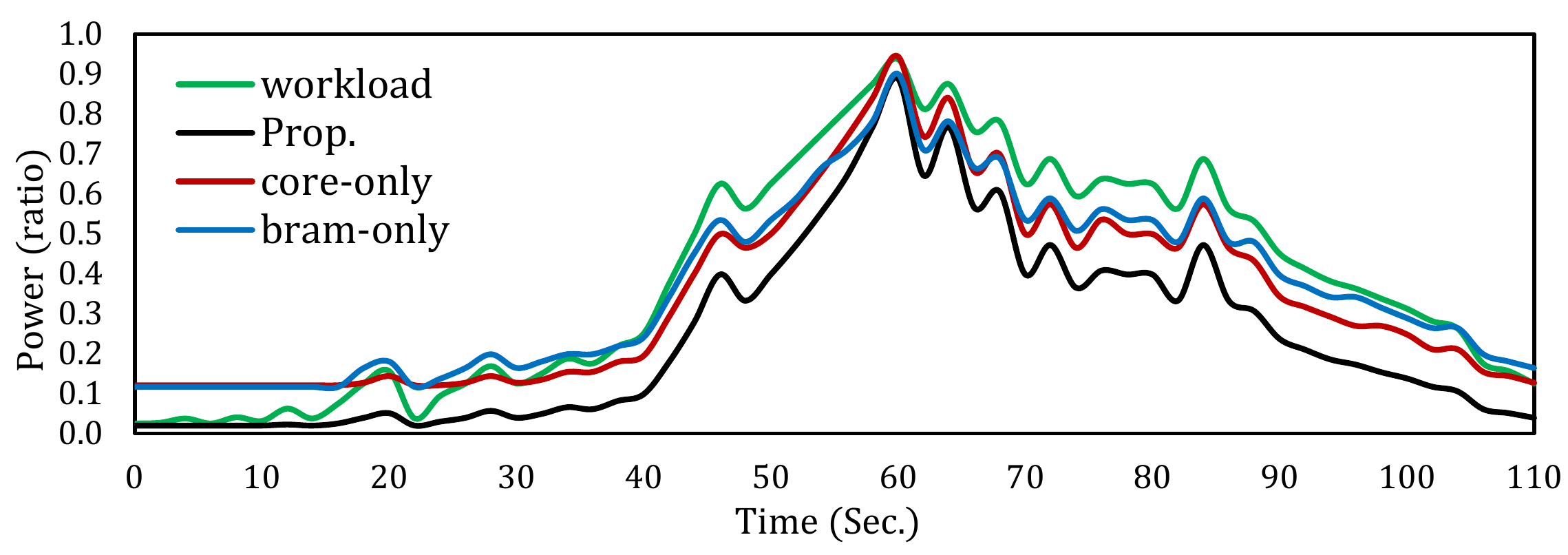}
  \vspace{-0.6cm}
  \caption{Comparing the efficiency of different voltage scaling techniques under a varying workload for Tabla framework.}  \label{fig:res-tabla1}
  \vspace{-0.5cm}
\end{figure}

\begin{figure}[t]
  \centering
  \includegraphics[width=0.5\textwidth]{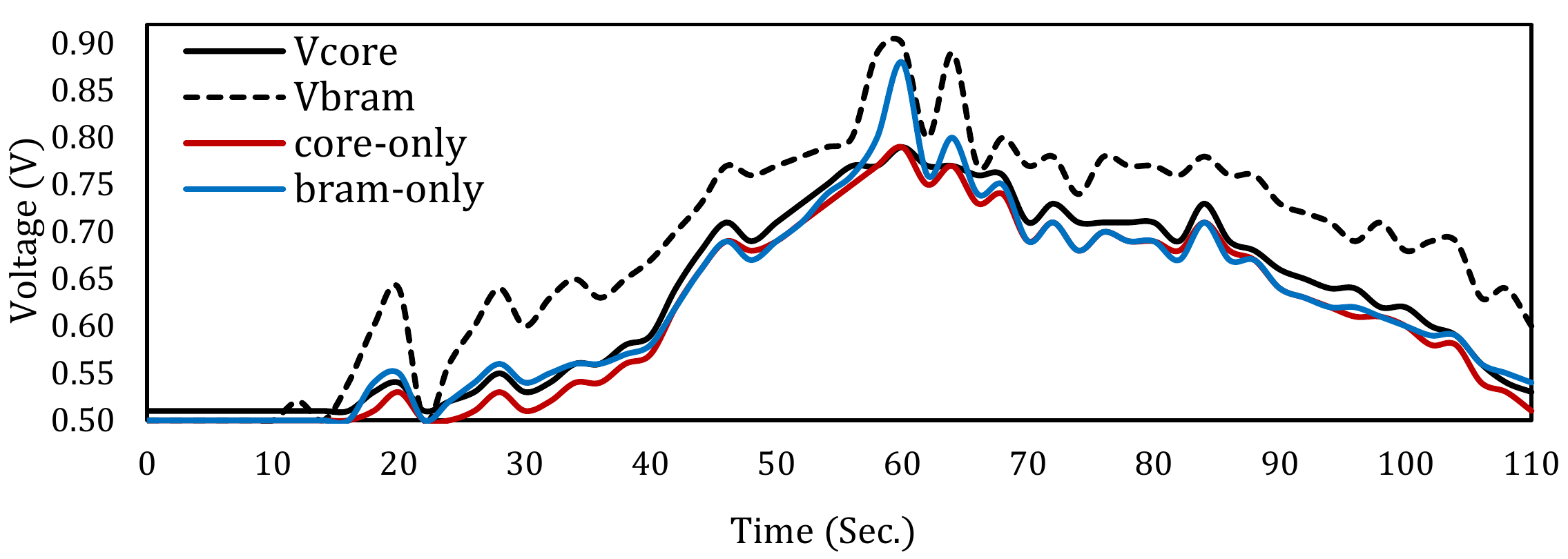}
  \vspace{-0.2cm}
  \caption{Voltage adjustment in different voltage scaling techniques under the varying workload for Tabla framework.}  \label{fig:res-tabla2}
  \vspace{-0.5cm}
\end{figure}

\subsection{Results}

Figure \ref{fig:res-tabla1} compares the achieved power gain of different voltage scaling approaches implemented the Tabla acceleration framework under a varying workload.
We considered a synthetic workload with 40\% average load (of the maximum) from \cite{yin2015burse} with $\lambda = 1000$, $H = 0.76$ and $IDC = 500$ where $\lambda$, $0.5 < H \leq 1$ and IDC denote the average arrival rate of the whole process, Hurst exponent, and the index of dispersion, respectively.
The workload also has been shown in the same figure (in green line) which is normalized to its expected peak load.
We have showed the corresponding $V_{core}$ and $V_{bram}$ voltages of all approaches in Figure \ref{fig:res-tabla2}.
Note that we have not showed $V_{bram}$ ($V_{core}$) for the core-only (bram-only) techniques as it is fixed 0.95V (0.8V) in this approach.
An average of $4.1\times$ power reduction is achieved, while this is $2.9\times$ and $2.7\times$ for the core-only and bram-only approaches.
This means that the proposed technique is 41\% more efficient than the best approach, i.e., only considering the core voltage rails.
An interesting point in Figure \ref{fig:res-tabla2} is the reaction of bram-only approach with respect to workload variation.
It follows a similar scaling trend (i.e., slope) as $V_{bram}$ in our approach.
However, our method also scales the $V_{core}$ to find more efficient energy point, thus $V_{bram}$ in our proposed approach is always greater than that of bram-only approach.

\begin{figure}[t]
  \centering
  \includegraphics[width=0.5\textwidth]{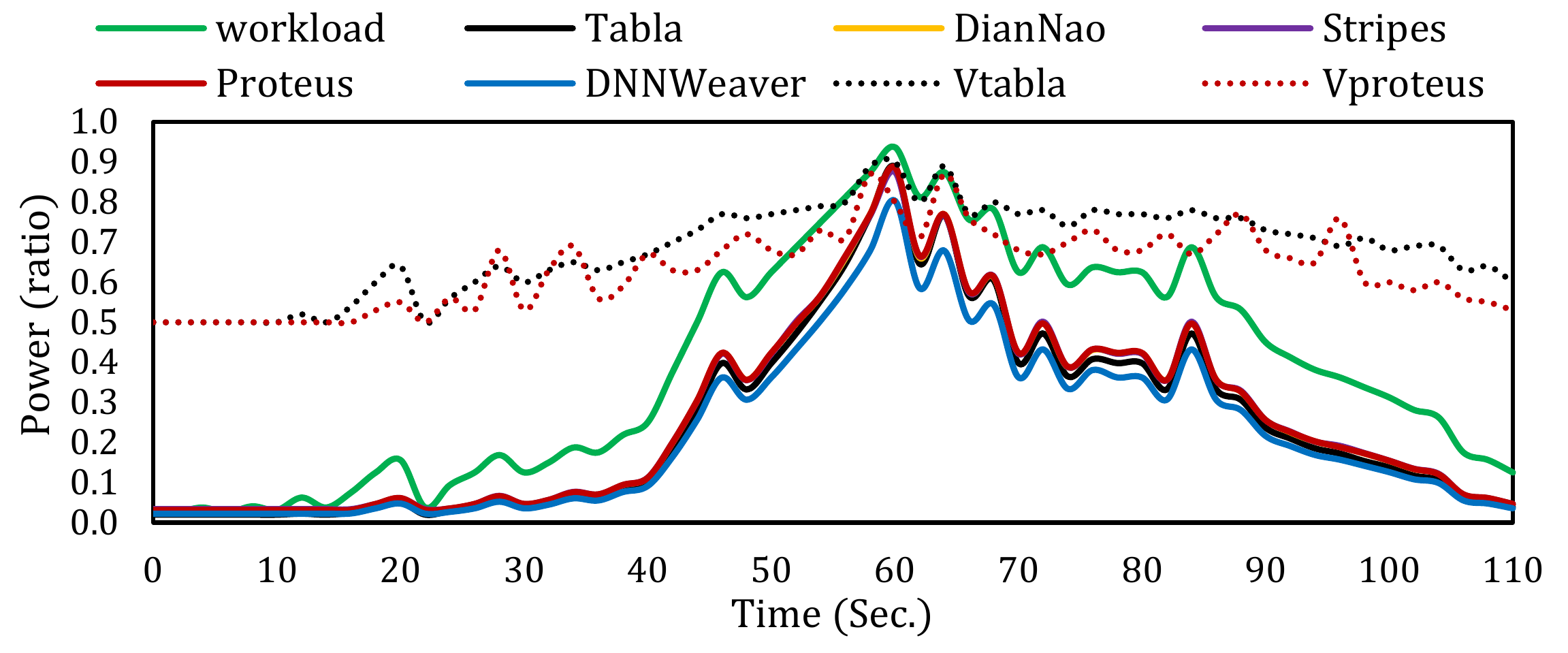}
  \vspace{-0.4cm}
  \caption{Power efficiency of the proposed technique in different acceleration frameworks.}  \label{fig:res-prop}
  \vspace{-0.3cm}
\end{figure}

Figure \ref{fig:res-prop} compares the power saving of all accelerator frameworks employing our proposed method, where they follow a similar trend.
This is due to the fact that the workload has considerably higher impact on the opportunity of power saving.
We could also infer this from Figure \ref{fig:P-W} where the power efficiency is significantly affected by workload load rather than the application specifications ($\alpha$ and $\beta$ parameters).
In addition, we observed that BRAM delay contributes to a similar portion of critical path delay in all of our accelerators (i.e., $\alpha$ parameters are close).
Lastly, the accelerators are heavily I/O-bound which are obliged to be mapped to a considerably larger device where static power of the unused resources is large enough to cover the difference in applications power characteristics.
Nevertheless, we have also represented the BRAM voltages of the Table ($V_{Tabla}$ in dashed black line, the same presented in Figure \ref{fig:res-tabla2}) and Proteus ($V_{Proteus}$) applications in \ref{fig:res-prop}. As we can see, although the power trends of these applications almost overlap, they have a noticeably different minimum $V_{bram}$ points.

\begin{table}[!t]
\caption{Comparison of power efficiency of different approaches.}
\vspace{-0.2cm}
\resizebox{0.5\textwidth}{!}{
\begin{tabular}{lllllll}
\hline
\textbf{Technique}  & \textbf{Tabla} & \textbf{DianNao} & \textbf{Stripes} & \textbf{Proteus} & \textbf{DNNWeav.} & \textbf{Average} \\ \hline
{Core-only}    & 2.9$\times$   & 3.1$\times$     & 3.1$\times$     & 3.1$\times$     & 2.9$\times$ & 3.02$\times$       \\ 
{Bram-only}    & 2.7$\times$   & 1.9$\times$     & 1.8$\times$     & 2.0$\times$     & 2.9$\times$ & 2.26$\times$       \\ 
\textit{The proposed} & 4.1$\times$   & 3.9$\times$     & 3.9$\times$     & 3.8$\times$     & 4.4$\times$ & 4.02$\times$       \\ 
\textit{Efficiency}  & 41-52\%      &   26-105\%      & 26-116\%        & 23-90\%        & 52\%     & 33.6\% $-$ 83\%      \\ \hline
\end{tabular} \label{tab:results}
}
\vspace{-0.5cm}
\end{table}

Table \ref{tab:results} summarizes the average power reduction of different voltage scaling schemes over the aforementioned workload.
On average, the proposed scheme reduces the power by 4.0$\times$, which is 33.6\% better than the previous core-only and 83\% more effective than scaling the $V_{bram}$.
As elaborated in Section \ref{sec:motivation}, different power saving in applications (while having the same workload) arises from different factors including the distribution of resources in their critical path where each resource exhibits a different voltage-delay characteristics, as well as the relative utilization of logic/routing and memory resources that affect the optimum point in each approach.

\section{Conclusion}
In this paper, we proposed an efficient framework to throttle the power consumption of multi-FPGA platforms by effectively scaling the voltage and frequency during runtime.
We utilize a light-weight predictor for proactive estimation of the incoming workload and incorporate it to our power-aware timing analysis framework that adjusts the frequency and finds optimal voltages according to the available workload margin, while maintaining the desired quality of service.
We evaluated the efficiency of our framework by implementing the state-of-the-art deep neural network accelerators on a solid FPGA architecture. Experimental results signified the efficiency of the proposed method, where we observed 4.0$\times$ power improvement, which is 33.6\% to 83\% more effective than previous approaches that merely consider a single voltage rail.

\section*{Acknowledgements}
This work was partially supported by CRISP, one of six centers in JUMP, an SRC program sponsored by DARPA, and also NSF grants \#1527034, \#1730158, \#1911095, and \#1826967.
We thank Prof. Moshovos's group from University of Toronto for providing the source codes of Stripes and Proteus.

\bibliographystyle{ieeetr}
\bibliography{mybibliography}

\end{document}